
\magnification=1200
\def\np#1#2#3{19#2 Nucl.Phys.B {\bf #1} #3.}
\def\pl#1#2#3{19#2 Phys.Lett. {\bf #1B} #3.}
\def\prd#1#2#3{19#2 Phys.Rev.D {\bf #1} #3.}
\def\prl#1#2#3{19#2 Phys.Rev.Lett. {\bf #1} #3.}
\def\cmp#1#2#3{19#2 Comm.Math.Phys. {\bf #1} #3.}
\def\mpl#1#2#3{19#2 Mod.Phys.Lett.A {\bf #1} #3.}

\def\jmp#1#2#3{19#2 J.Math.Phys. {\bf #1} #3.}
\line{\hfil IFUSP-P/1124}
\line{\hfil hep-th/9410003}
\vskip 1cm
\centerline{\bf TOPOLOGICAL DILATONIC SUPERGRAVITY THEORIES}
\vskip 1.5cm
\centerline{{\bf M.M.Leite  and {V.O.Rivelles}}{\footnote{*}{Partially
supported by CNPq}} }
\centerline{\it Instituto de F\'\i sica, Universidade de S\~ao Paulo}
\centerline{\it C. Postal 20516, CEP 01498-970, S\~ao Paulo, SP, Brazil}
\vskip 1.5cm

\centerline{October, 1994}

\vskip 4cm

\centerline{\bf Abstract}

We present a central extension of the $(m,n)$ super-Poincar\'e algebra
in two dimensions. Besides the usual Poincar\'e generators and the
$(m,n)$ supersymmetry generators we have $(m,n)$ Grassmann generators,
a bosonic internal symmetry generator and a central charge. We then
build up the topological gauge theory associated to this algebra.
We can  solve the classical field
equations for the fields which do not belong to the supergravity multiplet
and to a Lagrange multiplier multiplet.
The resulting topological supergravity theory turns out to be
non-local in the fermionic sector.

\hfill \eject

\bigskip \bigskip

{\bf 1.INTRODUCTION}

\bigskip

There is a renewed interest in the study of two dimensional gravity theories.
It stems from the fact that string theory predicts corrections to the usual
gravity theories. To lowest order (Callan 1985) these corrections need the
introduction
of a massless dilaton field and a massless antisymmetric tensor. We can then
ask what are the modifications introduced by these new fields with respect
to the results obtained from Einstein theory. In fact we would like to
know whether the conceptual and technical problems met in Einstein theory
can be overcome in this new framework. Of particular importance is
the process of black hole evaporation (Hawking 1975). In this case it is
possible to throw away the technical difficulties of the four dimensional
case while retaining the essential features responsible for the back reaction
if we consider a simplified version of the model in two
dimensions (Callan 1992).
When we consider only the graviton and the dilaton fields in this simplified
model it turns out that the resulting theory is a topological gravity
theory (Verlinde 1991). This topological theory is obtained by gauging
the Poincar\'e algebra in two dimensions. However in this formulation
the fields
have unusual transformation properties under Poincar\'e gauge transformations
(Verlinde 1992).
This can be overcome if we take as the gauge algebra a central extension
of the Poincar\'e algebra where the usual commutator of the translation
generators $P_a$ is replaced by $[P_a, P_b] = \epsilon _{ab} Z$ where
$Z$ is a central charge (Cangemi 1992).  This central extension of the
Poincar\'e algebra has also been considered in the context of WZW models.
It was used as a prototype for WZW models
with non-semisimple groups. It is remarkable that the corresponding sigma
model describes string propagation on a four dimensional gravitational
plane wave background (Nappi 1993).

Supersymmetry was introduced in dilaton gravity theories in two
dimensions with the
purpose of studying the positive energy theorem (Park 1993) and black
holes (Nojiri 1993). These formulations are the naive supersymmetric
extensions of the bosonic theories to a superspace and are not topological.
Supergravity was formulated as a topological theory by
gauging the groups $OSp(2 \mid 1)$ and $OSp( 2 \mid 2)$ and the
resulting theory was respectively the $N = 1$ and $N = 2$
super de Sitter theories
(Montano 1990, Li 1990). Only recently the $(0,1)$ (Cangemi 1994)
and $(1,1)$ (Rivelles 1994) topological supergravity models were
obtained. For the $(0,1)$ case it was only needed the introduction
of the supersymmetry generators for the centrally extended Poincar\'e
algebra while for the $(1,1)$ case besides the supersymmetry
generators it was needed the introduction of another Grassmann generator
(which does not generate supersymmetry) and a new bosonic generator (besides
the central charge) (Rivelles 1994). Remarkably most of the resulting field
equations can be solved. If we keep only the supergravity fields and a
supermultiplet of Lagrange multipliers it can be shown (Rivelles 1994) that
by eliminating the extra gauge and  Lagrange multiplier fields the
resulting theory is a supergravity theory non-local in the fermionic
sector. Of course the non-local terms can be removed by the introduction
of an auxiliary spinor field.

In this paper we present the full extended supersymmetric version of the
dilaton gravity theories. We start by presenting the central extension
of the $(m,n)$ super-Poincar\'e algebra. We then build
up the quadratic Casimir operator from which a non-degenerated metric on
the group can be obtained. The gauge theory for the $(m,n)$ algebra is then
written and the proper field equations are solved to obtain the
corresponding extended supergravity theory. As in the cases $(0,1)$
and $(1,1)$ the resulting supergravity theory turns out
to be non-local in the fermionic sector.

The plan of the paper is as follows: in Section 2 we present the
centrally extended $(m,n)$ super-Poincar\'e algebra. In section 3
in order to have a self contained presentation we review the topological
formulation of the central extension of the Poincar\'e algebra. Finally
in Section 4 we present the $(m,n)$ topological supergravity theory
based on the gauging of the algebra obtained in Section 1. We also
solve some of the field equations and discuss the gauge fixing needed
to obtain a supergravity theory involving only the supergravity fields and
a multiplet of Lagrange multipliers. In Section  4 we comment on some
possible applications of the $(m,n)$ algebra.

\bigskip \bigskip

{\bf 2. CENTRAL EXTENSION OF THE $(m,n)$ SUPER-POINCAR\'E ALGEBRA}

\bigskip

The central extension of the Poincar\'e algebra, which is peculiar to
two dimensions, is given by (Cangemi 1992)

$$ [P_a, P_b] = \epsilon_{ab} Z $$
$$[J, P_a] = {\epsilon_a}^b P_b $$
$$[P_a, Z] = [J, Z] = 0
\eqno (2.1) $$
where $P_a$ is the translation generator, $J$ is the Lorentz transformation
generator and $Z$ is a central charge. The flat Minkowski metric is $h_{ab} =
diag(-1, +1)$ and $\epsilon^{01} = 1$. This algebra can be obtained from the
De Sitter algebra $SO(2,1)$ by an unconventional contraction where
$J$ is replaced by $J + {Z\over\lambda}$ and then the limit $\lambda
\rightarrow \infty$ is taken (Cangemi 1992). It was also obtained earlier as
a generalization of the In\"on\"u-Wigner contraction (Saletan 1961).

The central extension of the $(m,n)$ super Poincar\'e algebra requires the
introduction of the  Grassmann generators
$Q^{+i}, U^{-i} (i = 1 \dots m)$, and $ Q^{-I}, U^{+I} (I = 1 \dots n)$,
and the bosonic generators $K^{iJ}$. Here $Q$ is the generator of
supersymmetry. They close the following algebra together with Eq.(2.1)

$$ [{ J},{ Q}_{\alpha}^{+i} ]=-{1\over2} { Q}_{\alpha}^{+i}, \,\,\,
   [{ J},{ Q}_{\alpha}^{-I} ]= {1\over2} { Q}_{\alpha}^{-I} $$
$$ [{ J},{ U}_{\alpha}^{+I}]= -{1\over2} { U}_{\alpha}^{+I}, \,\,\,
   [{ J},{ U}_{\alpha}^{-i}]= {1\over2} { U}_{\alpha}^{-i} $$
$$ [{ P}_a,{ Q}_{\alpha}^{+i} ] ={1\over2}
(\gamma_a{ U}^{-i})_{\alpha} $$
$$ [{ P}_a,{ Q}_{\alpha}^{-I} ]= -{1\over2}
(\gamma_a { U}^{+I})_{\alpha} $$
$$ [ { K}^{iJ},{ Q}_{\alpha}^{+k} ]=
-{1\over2}{\delta}^{ik}{ U}_{\alpha}^{+J} $$
$$ [ { K}^{iJ},{ Q}_{\alpha}^{-K} ]=
-{1\over2}{\delta}^{JK}{ U}_{\alpha}^{-i} $$
$$\{{ Q}_{\alpha}^{+i},{ Q}_{\beta}^{+j} \}=
(\gamma_{+}\gamma^{a}C)_{\alpha\beta}{\delta}^{ij}{ P}_a $$
$$ \{{ Q}_{\alpha}^{-I},{ Q}_{\beta}^{-J} \}=
-(\gamma_{-}\gamma^{a}C)_{\alpha\beta}{\delta}^{IJ}{ P}_a $$
$$ \{{ Q}_{\alpha}^{+i},{ Q}_{\beta}^{-J} \}=
(\gamma_{+}C)_{\alpha\beta}{ K}^{iJ} $$
$$ \{{ Q}_{\alpha}^{+i},{ U}_{\beta}^{-j} \}=
-(\gamma_{+}C)_{\alpha\beta}{\delta}^{ij} { Z} $$
$$ \{{ Q}_{\alpha}^{-I},{ U}_{\beta}^{+J} \}=
(\gamma_{-}C)_{\alpha\beta}{\delta}^{IJ} { Z} $$
$$ [{ P}_a,{ K}^{iJ} ] = [{ J},{ K}^{iJ} ]=0 $$
$$ [ { P}_a ,{ U}_{\alpha}^{+I} ]=[ { P}_a ,
{ U}_{\alpha}^{-i} ]=\{{ Q}_{\alpha}^{+i},{ U}_{\beta}^{+J} \}=
\{{ Q}_{\alpha}^{-I},{ U}_{\beta}^{-j} \}=0 $$
$$ \{{ U}_{\alpha}^{+I},{ U}_{\beta}^{+J} \}=\{{ U}_{\alpha}^{+I},
{U}_{\beta}^{-j} \}=\{{ U}_{\alpha}^{-i},{ U}_{\beta}^{-j} \}=0
\eqno (2.2)$$
Our conventions are the following:  the Dirac gamma matrices are
$\gamma^0 = i \sigma_2, \gamma^1 =
\sigma_1, \gamma_2 = \sigma_3$ and $C = i \sigma_2$ where $\sigma_i$ are
the Pauli matrices;  the chiral projections are defined as
$\psi^\pm = {1\over2}( 1 \pm \gamma_2 ) \psi$. The algebra Eq.(2.2) satisfy
the graded Jacobi identities.

Previously only the $(0,1)$ and the $(1,1)$ cases were known.
The minimal supersymmetric version of the extended Poincar\'e algebra
consists in adding just one chiral supersymmetry generator.
It is denoted by $(0,1)$ or $(1,0)$ algebra depending on the
chirality chosen.
In  (Cangemi 1994) the fermionic generators are $Q^-$ for supersymmetry and
$U^+$ for the extra fermionic charge. The fermionic generators
satisfy the following (anti-)commutation relations besides the ones in
Eq.(2.1)
\footnote{*}{The relation between the fermionic generators of (Cangemi 1994)
and ours is just $Q^-_{(ours)} = Q^-_{(Cangemi)}, U^+_{(ours)} =
16 i Q^+_{(Cangemi)}, P_{a {(ours)}} = 4 i P_{a {(Cangemi)}}$ and
$Z_{(ours)} = 16 Z_{(Cangemi)}$ besides the opposite convention for the
sign of $\epsilon^{ab}$. }

$$[J, Q^-_\alpha] = {1\over2} Q^-_\alpha, \,\,\,
  [J, U^+_\alpha] = - {1\over2} U^+_\alpha $$
$$ [P_a, Q^-_\alpha] = - {1\over2} ( \gamma_a U^+)_\alpha $$
$$\{Q^-_\alpha, Q^-_\beta \} = - ( \gamma_- \gamma^a C )_{\alpha \beta}
P_a  $$
$$\{Q^-_\alpha, U^+_\beta \} = ( \gamma_- C )_{\alpha \beta} Z
\eqno (2.3)$$

When both chiral supersymmetry generators $Q^+$ and $Q^-$ are present we
can replace them by a single Majorana generator $Q$. In this case the
full algebra needs the introduction of one more bosonic generator $K$.
We also need one more Grassmann generator
$U^-$ besides $U^+$ and like $Q^\pm$ they are replaced by a single Majorana
generator $U$. They close the $(1,1)$ algebra (Rivelles 1994)

\vfill \eject
$$[J, Q_\alpha] = - {1\over2} ( \gamma_2 Q)_\alpha, \,\,\,
  [J, U_\alpha] = - {1\over2} ( \gamma_2 U)_\alpha $$
$$ [P_a, Q_\alpha] = {1\over2} ( \gamma_a U)_\alpha $$
$$[K, Q_\alpha] = - {1\over2} ( \gamma_2 U )_\alpha $$
$$\{Q_\alpha, Q_\beta \} = ( \gamma^a C )_{\alpha \beta} P_a - ( \gamma_2 C
)_{\alpha \beta} K $$
$$\{Q_\alpha, U_\beta \} = - ( \gamma_2 C )_{\alpha \beta} Z
\eqno (2.4) $$

Both algebras Eqs.(2.3) and (2.4) can be obtained by an unconventional
contraction of the super De Sitter algebra in a very similar way as the
algebra Eq.(2.1) was obtained from de De Sitter group (Rivelles 1994).

\bigskip \bigskip

{\bf 3. TOPOLOGICAL GRAVITY THEORIES}

\bigskip

In this section we will formulate the topological gravity theory on the
algebra Eq.(2.1). For simplicity we will consider only the bosonic case. The
topological supergravity theory will be discussed in the next section.

In two dimensions a topological theory has the action

$$ S = \int Tr ( \eta F )
\eqno (3.1)$$
where $F$ is the two-form field strength $F = dA + A^2$,  $A$ is the
one-form gauge potential and $\eta$ is a zero-form  Lagrange
multiplier in the coadjoint representation of the algebra. The action
Eq.(3.1) is invariant under the usual gauge transformations
$$ \delta A^a = D \theta^a  \equiv d \theta^a + f_{bc}^a A^b \theta^c $$
$$ \delta \eta_a = - f_{ab}^c \eta_c \theta^b
\eqno (3.2) $$
where $\theta$ is the gauge parameter. The field
equations which follow from the action Eq.(3.1) are just $F = 0$ and
$ D \eta = 0$ where $D$ is the appropriate covariant derivative on the
algebra as defined in Eq.(3.2).

When we consider the algebra Eq.(2.1) we can expand the gauge potential
in terms of the generators of the algebra

$$ A = e^a P_a + w J  + A Z
\eqno(3.3) $$
The fields $e^a, w$ and $A$ are going to be identified with the
zweibein, the spin connection and an abelian gauge field, respectively.
The Lagrange multiplier $\eta$ can be expanded as

$$ \eta = \eta^a P_a + \eta J + \eta^{\prime} Z
\eqno (3.4) $$
where $\eta^a$ will turn out to be an auxiliary field, $\eta$ the
dilaton and $\eta^{\prime}$ the cosmological constant.

In order to write the action Eq.(3.1) we need to know the metric on
the algebra. It can be read off from the quadratic Casimir operator.
For the algebra Eq.(2.1) the quadratic Casimir operator is

$$ C^{(2)} = P^a P_a + Z J + J Z
\eqno (3.5) $$
{}from which we get the nondegenerated metric

$$<P_a, P_b> = h_{ab},\,\,\, <J, Z> = 1
\eqno(3.6) $$
Then the action Eq.(3.1) can be written as

$$ S = \int [ \eta_a F^a(P) + \eta F(J) +  \eta^{\prime} F(Z) ]
\eqno(3.7) $$

The field equations which follow from $F = 0$ are just
$$  D e^a  \equiv de^{a} + we^{b}\epsilon_{b}^{a} = 0
\eqno (3.8a) $$
$$  dw = 0
\eqno (3.8b) $$
$$ dA + {1\over2}e^{a}e^{b}\epsilon_{ab} = 0
\eqno (3.8c) $$
and those following from the variation of the gauge potential, i.e.
$D\eta = 0$, are
$$ D \eta_{a} + \eta^{\prime} \epsilon_{ab} e^b = 0
\eqno (3.9a) $$
$$ d\eta + e^{b} \epsilon_{b}^{\;\;a} \eta _{a} = 0
\eqno (3.9b) $$
$$   d\eta^{\prime}=0
\eqno (3.9c) $$
Notice that some field equations are just algebraic equations. Then Eq.(3.8a)
can be solved for $w$ and gives the usual expression for the spin connection.
Eq.(3.9b) can be solved for $\eta_a$. The field equation for $\eta^\prime$
Eq.(3.9c) is a first order differential equation which can be solved
locally and it states that $\eta^\prime$ is a
constant which we are going to identify with the cosmological
constant $\Lambda$. After
substituting these solutions of the field equations back into the action
Eq.(3.7) we obtain the usual form for the dilaton gravity action
$$  S = \int ( \eta R + {\Lambda\over2} e^a e^b \epsilon_{ab} )
\eqno (3.10) $$
where $R$ is the curvature scalar. The action Eq.(3.10) should be
complemented by the field equation Eq.(3.8c) for the abelian gauge
field $A$. Eq.(3.8c) can be solved once
we have a solution for the gravitational sector.

\bigskip \bigskip

\vfill \eject

{\bf 4. $(m,n)$ TOPOLOGICAL DILATONIC SUPERGRAVITY }

\bigskip

We can now proceed in an analogous way to obtain the $(m,n)$ topological
dilatonic supergravity theory
using the central extension of the $(m,n)$ super-Poincar\'e algebra Eq.(2.2).
We first find the quadratic Casimir operator
$$ C^{(2)} = P_{a}P^{a} - K^{iJ}K^{iJ} + JZ + ZJ +
{1\over2} C^{\alpha\beta} ( Q_{\alpha}^{+i}U_{\beta}^{-i} +
U_{\alpha}^{-i}Q_{\beta}^{+i} + U_{\alpha}^{+I}Q_{\beta}^{-I} +
Q_{\alpha}^{-I}U_{\beta}^{+I} )
\eqno (4.1) $$
{}from which we can read off the nondegenerated metric
$$  < P_{a},P_{b} > = h_{ab}, \,\,\,
  < K^{iI},K^{jJ} > = - \delta^{ij}\delta^{IJ}, \,\,\,
  < J,Z >= 1, $$
$$  < Q_{\alpha}^{+i},U_{\beta}^{-j} > =
{1\over2}C_{\alpha\beta} \delta^{ij}, \,\,\,
  < U_{\alpha}^{-i},Q_{\beta}^{+j} > =
{1\over2}C_{\alpha\beta} \delta^{ij}, $$
$$  < U_{\alpha}^{+I},Q_{\beta}^{-J} > =
{1\over2}C_{\alpha\beta} \delta^{IJ},  \,\,\,
  < Q_{\alpha}^{-I},U_{\beta}^{+J} > =
{1\over2}C_{\alpha\beta} \delta^{IJ}
\eqno (4.2) $$

The gauge potential $A$ has  the following expansion
$$ A = e^{a}P_{a} + wJ + v^{iJ}K^{iJ} + A Z +
\psi^{\alpha +i} Q_{\alpha}^{+i} +
\psi^{\alpha -I} Q_{\alpha}^{-I} +
{\xi}^{\alpha +I} U_{\alpha}^{+I} +
\xi^{\alpha -i} U_{\alpha}^{-i}
\eqno (4.3) $$
and the field strength
$$F = F^{a}(P)P_{a} + F(J)J + F^{iJ}(K)K^{iJ} + F(Z) Z + $$
$$ + F^{\alpha i}(Q^{+})Q_{\alpha}^{+i} + F^{\alpha I}(Q^{-})Q_{\alpha}^{-I} +
F^{\alpha I}(U^{+})U_{\alpha}^{+I} + F^{\alpha i}(U^{-})U_{\alpha}^{-i}
\eqno (4.4) $$
has the following components

$$ F^{a}(P) = De^{a} + {1\over2}\psi^{-I}\gamma^{a}
\psi^{+I} - {1\over2}\psi^{+i}\gamma^{a}\psi^{-i}
\eqno (4.5a) $$
$$ F(J) = dw \eqno (4.5b) $$
$$ F^{\alpha i}(Q^{+}) = D \psi^{\alpha +i} \equiv d \psi^{\alpha +i} -
{1\over2} w \psi^{\alpha +i}
\eqno (4.5c) $$
$$ F^{\alpha I}(Q^{-}) = D \psi^{\alpha -I} \equiv d\psi^{\alpha -I} +
{1\over2} w \psi^{\alpha -I}
\eqno (4.5d) $$
$$ F^{\alpha I}(U^{+}) = D \xi^{\alpha +I} -
{1\over2}e^{a}(\psi^{-I}\gamma_{a})^{\alpha}
- {1\over2} v^{iI} \psi^{\alpha +i}
\eqno (4.5e) $$
$$ F^{\alpha i}(U^{-}) = D \xi^{\alpha -i} +
{1\over2}e^{a}(\psi^{+i}\gamma_{a})^{\alpha}
 + {1\over2} v^{iI} \psi^{\alpha -I}
\eqno (4.5f) $$
$$ F^{iJ}(K) = dv^{iJ} + \psi^{+i}\psi^{+J}
\eqno (4.5g) $$
$$ F(Z) = dA + {1\over2}e^{a}e^{b}\epsilon_{ab} +
\psi^{+i}\xi^{+i} - \psi^{-I}\xi^{-I}
\eqno (4.5h) $$
The Lagrange multiplier $\eta$ can be expanded as
$$\eta = \eta^{a}P_{a} +
\eta J  + \eta^{iJ}K^{iJ} + \eta^{\prime \prime}Z
+ \chi^{\alpha -i}Q_{\alpha}^{+i} +
  \chi^{\alpha +I}Q_{\alpha}^{-I} +
  \zeta^{\beta -I}U_{\beta}^{+I} +
  \zeta^{\beta +i}U_{\beta}^{-i}
\eqno (4.6) $$
and the action takes the form
$$     S = \int [\eta _{a} F^{a}(P) + \eta F(J) + \eta^{ iJ} F(K^{iJ}) +
 \eta^{\prime \prime} F(Z) + $$
$$ +{\chi}_{\alpha}^{+i} F^{\alpha +i}(Q)
   + {\chi}_{\alpha}^{-I} F^{\alpha -I}(Q)
   + {\zeta}_{\alpha}^{+I} F^{\alpha +I}(U)
   + \zeta^{ -i}_{\alpha} F^{\alpha -i}(U)]
\eqno (4.7) $$

As in the pure bosonic case we can solve some of the field equations and
remain only with the supergravity fields $e^a$ and $\psi^\pm$ and a
multiplet of Lagrange multipliers $\eta$ and $\chi^\pm$. By varying the
action Eq.(4.7) with respect to the gauge fields we obtain

$$ D \eta_{a} + \eta^{\prime \prime} \epsilon_{ab} e^{b}
 + {1\over2} \psi^{-I} \gamma_{a} \zeta^{+I} - {1\over2} \psi^{+i}
\gamma_{a} \zeta^{-i} = 0
\eqno (4.8a) $$
$$ d\eta + e^{b} \epsilon_{b}^{\;\;a} \eta _{a} +
{1\over2}\psi^{+i} \chi^{+i} -{1\over2}\psi^{ -I} \chi^{ -I} +
{1\over2} \xi^{+I} \zeta^{+I} -{1\over2} \xi^{-i} \zeta^{ -i}
= 0
\eqno (4.8b) $$
$$  d\eta^{iJ} + {1\over2} \psi^{+i} \zeta^{+J} -
{1\over2}\psi^{ -J}\zeta^{-i} =0
\eqno (4.8c) $$
$$   d\eta^{\prime \prime}=0
\eqno (4.8d) $$
$$ D \chi_{\alpha}^{+i} +
\eta_{a}(\gamma^{a} \psi^{-i})_{\alpha}
- \eta^{\prime \prime} \xi_{\alpha}^{ +i} -
{1\over2} e^{a}(\gamma_{a} \zeta^{ -i})_{\alpha}
 + {1\over2}v^{iJ} \zeta_{\alpha}^{+J} +
\eta^{iJ} \psi_{\alpha}^{ +J} = 0
\eqno (4.8e) $$
$$ D \chi_{\alpha}^{ -I} -
\eta_{a}(\gamma^{a} \psi^{ +I})_{\alpha} +
\eta^{\prime \prime} \xi_{\alpha}^{-I}
+ {1\over2} e^{a}(\gamma_{a} \zeta^{+I})_{\alpha} -
{1\over2} v^{iI} \zeta_{\alpha}^{ -i}
 + \eta^{jI} \psi_{\alpha}^{-j} = 0
\eqno (4.8f) $$
$$  D \zeta_{\alpha}^{+I} -
\eta^{\prime \prime}\psi_{\alpha}^{ +I} = 0
\eqno (4.8g) $$
$$ D \zeta_{\alpha}^{ -i}
  + \eta^{\prime \prime} \psi_{\alpha}^{-i} = 0
\eqno (4.8h) $$

Some of the field equations for the gauge fields can also be solved. Setting
$F(P) = 0$ in Eq.(4.5a) we can solve algebraically for $w$ obtaining the
usual supersymmetric spin connection
$$  w = -(det\;e)^{-1} e^{a} \epsilon^{\mu\nu}  ( \partial_{\mu}
e_{\nu}^{b} h_{ab}
-{1\over2} \psi_{\mu}^{+i} \gamma_{a} \psi_{\nu}^{-i} + {1\over2}
 \psi_{\mu}^{-I}
 \gamma_{a} \psi_{\nu}^{+I} )
\eqno (4.9) $$
Then Eqs.(4.5e, f, g) can be solved for $\xi^{+J}, \xi^{ -i}$ and
$ v^{iJ}$,
$$ v^{iJ} = - {1\over{2\eta^{\prime \prime}}}(\psi^{+i}\zeta^{+J} +
\psi^{-J} \zeta^ { -i})$$
$$ \xi^{\alpha +J}= -{1\over{ 2\eta^{\prime \prime}}} e^{a}(\zeta^{-J}
\gamma_{a})^{\alpha} - {1\over{(4 \eta^{\prime \prime})}^2}
\zeta^{-J} \zeta^{ -i} \psi^{\alpha +i}$$
$$  \xi^{\alpha -i}=
-{1\over{2\eta^{\prime \prime}}}e^{a}(\zeta^{ +i}
\gamma_{a})^{\alpha} - {1\over4(\eta^{\prime \prime})^{2}}
\zeta^{ +i} \zeta^{+J} \psi^{\alpha -J}
\eqno (4.10) $$

{}From the field equations for the Lagrange multipliers Eqs.(4.8a,c,d)
we can solve for $\eta_a, \eta^{iJ}$ and for $\eta^{\prime \prime}$.
The solution for $\eta^{\prime \prime}$ is just $\eta^{\prime \prime} =
constant$ which can be later identified with the cosmological constant, and
$$ \eta_a = \epsilon_a^b e_b^\mu ( - \partial_\mu \eta -
{1\over2} \psi_\mu^{+i} \chi^{+i} +
{1\over2} \psi_\mu^{-I} \chi^{ -I} -
{1\over2} \xi_\mu^{+I} \zeta^{+I} +
{1\over2} \xi_\mu^{ -i} \zeta^{ -i}) $$
$$ \eta^{iJ} = {1\over{2 \eta^{\prime \prime}}} \zeta^{+i} \zeta^{+J}
\eqno(4.11) $$

By substituting the solutions Eqs.(4.9-11) into the action Eq.(4.7) we
obtain the effective action for the $(m,n)$ extended supergravity
$$  S = \int (  \eta dw + \chi_{\alpha}^{+i}D\psi^{\alpha +i} +
 \chi_{\alpha}^{ -I}D \psi^{\alpha -I}
 + {1\over2} \eta^{\prime \prime} e^{a}e^{b}\epsilon_{ab} + $$
$$ + {1\over2} e^{a} \psi^{-I}\gamma_{a} \zeta^{+I} -
 {1\over2} e^{a} \psi^{+i}\gamma_{a} \zeta^{-i}  +
 {1\over{2\eta^{\prime \prime}}} \zeta^{ +i}\zeta^{+J}
\psi^{-J} \psi^{-i}  )
\eqno (4.12) $$

In order to find out the supergravity transformations which leave the
action Eq.(4.12) invariant some care must be taken. Since we have solved
the field  equations for some of the gauge fields the gauge transformations
Eq.(3.2) associated to these gauge fields must be fixed. Taking the
solutions Eqs.(4.10) and (4.11)  for
$v, \eta$ and $\xi$ we can find that under a gauge
transformation Eq.(3.2) generated by $Q$ they transform with a supersymmetry
transformation with
parameter $\epsilon$ plus a gauge transformation generated by $\xi$ with
parameter $\epsilon^\prime = 0$ plus a gauge transformation generated
by $v$ with parameter $\alpha^{iJ} = - {1\over{2 \eta^{\prime \prime}}}
( \epsilon^{+i} \zeta^{+J} + \epsilon^{-J} \zeta^{-i} )$.
Since $\eta$ and $\chi$ have a non-trivial
gauge transformation generated by $v$  and $\xi$ the resulting supergravity
transformation is a sum of a supergravity transformation with parameter
$\epsilon$ plus a gauge transformation generated by $\xi$ with
parameter $\epsilon^\prime $ plus a gauge transformation generated
by $v$ with a parameter $\alpha $ with the values of $\epsilon^\prime $
and $\alpha $ given above. The supergravity transformation on the
remaining fields is the same as the ones generated by $Q$. We then find

$$ \delta e_{\mu}^{a} = \epsilon^{+i}\gamma^{a}\psi_{\mu}^{-i} -
\epsilon^{ -I}\gamma^{a}\psi_{\mu}^{+I} $$
$$ \delta \psi^{\alpha +i} = D \epsilon^{\alpha +i} $$
$$ \delta \psi^{\alpha -I} = D \epsilon^{\alpha -I} $$
$$ \delta \chi_\alpha^{+i} = - {1\over{2 \eta^{\prime\prime}}} \zeta^{+i}
\zeta^{+J} \epsilon^{+J}_\alpha  -
 \eta_a \gamma^a \epsilon^{-i}_\alpha $$
$$ \delta \chi_\alpha^{ -I} = - {1\over{2 \eta^{\prime\prime}}} \zeta^{+j}
\zeta^{+J} \epsilon^{-i}_\alpha  +
 \eta_a \gamma^a \epsilon^{+I}_\alpha $$
$$ \delta \zeta_\alpha^{ -i} =
-\eta^{\prime \prime} \epsilon^{-i}_\alpha $$
$$ \delta \zeta_\alpha^{+I} =
\eta^{\prime \prime} \epsilon^{ +I}_\alpha $$
$$  \delta \eta = - {1\over2}\epsilon^{+i}\chi^{+i} +
{1\over2}\epsilon^{ -I} \chi_\alpha^{ -I}
\eqno (4.13)$$
where $\eta_a$ is given by Eq.(4.11).
These transformations leave the action Eq.(4.12) invariant and close on-sehll
on a local translation, a local Lorentz transformation and a supergravity
transformation.

Notice that in order to have just a set of supergravity fields we should
eliminate $\zeta^\pm$ from the action Eq.(4.12). By doing that the action
would become non-local in the fermionic sector so we choose to keep the
auxiliary field $\zeta^\pm$. Then we have a new version for supergravity
in two dimensions which differ, in the case $(1,1)$, from the ones knew
before (Nojiri 1993, Park 1993) as discussed in (Rivelles 1994).

\bigskip \bigskip

{\bf 5. CONCLUSIONS }

\bigskip

We have presented the $(m,n)$ supersymmetric version of the extended
Poincar\'e algebra. In order to obtain it we had to introduce a new
Grassmann generator $U$ and a new bosonic generator $K$ besides the
supersymmetry generator $Q$. We then build up in the usual way the
corresponding supergravity theory. When we eliminated the extra gauge
and Lagrange multiplier fields the remaining supergravity fields
have an action which is non-local in the fermionic sector.

As the extended Poincar\'e algebra has been used as a prototype
for the study of non-semisimple WZW models the $(m,n)$ algebra
presented in this paper can be used as a prototype for the supersymmetric
case. It would be possible then to find out exact string backgrounds
which are supersymmetric.

Also the non-local supergravity model presented in Section 4 should
give rise to some new class of induced supergravity theories. Since the
non-locality appears only in the fermionic sector it would be very
interesting to find out the relation of this model to the know induced
supergravity models. Other aspects, like the relation to supersymmetric
matrix models, properties of black holes and the quantization
of the model remain to be investigated.

\bigskip \bigskip

{\bf Acknowledgments}

\bigskip

MML would like to thank FAPESP for financial support.

\vfill \eject

{\bf REFERENCES}

\bigskip \bigskip

Cangemi D and Jackiw R \prl{69}{92}{233}

Cangemi D and Leblanc M \np{420}{94}{363}

Callan C G, Friedan D, Martinec E J and Perry M J \np{262}{85}{593}

Callan C G, Giddings S B and  Harvey J A \prd{45}{92}{R1005}

Hawking S \cmp{43}{75}{199}

Li K \np{346}{90}{329}

Montano D, Aoki K and Sonnenschein J \pl{247}{90}{64}

Nappi C R and Witten E \prl{71}{93}{3751}

Nojiri S and Oda I \mpl{8}{93}{53}

Park Y and Strominger A \prd{47}{93}{1569}

Rivelles V O \pl{321}{94}{189}

Saletan E.J. \jmp{2}{61}{1}

Verlinde E and Verlinde  H \np{348}{91}{457}

Verlinde H 1992 {\it String Theory and Quantum Gravity '91}, Eds. Harvey J,
Iengo R,
\indent Narain K S and Verlinde H, p.178 (World Scientific, Singapore)

\vfill \eject
\end